\RequirePackage{fix-cm}
\documentclass[smallextended]{svjour3}       
\smartqed  
\usepackage[colorlinks,citecolor=blue,bookmarks]{hyperref}
\usepackage{graphicx}
\begin{document}

\title{A microscopically derived formula for alpha-decay half-lives}

\titlerunning{A microscopically derived formula for alpha-decay...}      

\author{Swagatika Bhoi, Basudeb Sahu*}

\institute{Swagatika Bhoi \at
              School of Physics, Sambalpur University, Jyoti Vihar, Burla,768019, India.\\
              \email{bd\_sahu@yahoo.com}           
           \and
           Basudeb Sahu*  \at
              Department of Physics, College of Engineering and Technology
Bhubaneswar-751003, India.
}

\date{Received: date / Accepted: date}

\maketitle

\begin{abstract}
Although Geiger-Nuttall (GN) law gives a single straight line, if we consider the experimental data of alpha particle emitters including heavy and super heavy nuclei with proton numbers as large as 118, instead of getting a single linear path we observe several linear segments with different slopes and intercepts. This problem is overcome when the experimental results in logarithm form are plotted as a function of Viola-Seaborg (VS) parameters with values of parameters set by hand. By using the fundamental principle of decay, we derive a formula of logarithm of half-lives in terms of well-defined parameters or coefficients and this replaces the empirical VS rule. 
\end{abstract} 

\section{Introduction}
The quintessential $\alpha$-radioactivity has been studied by many physicists so far and has opened doors for laying a rigid foundation and development of nuclear physics \cite{a,b,c,d}. Gamow was the first and foremost to apply quantum mechanics to a nuclear physics problem by providing the first model to explain $\alpha$-decay and propounded that the process involves tunneling of an $\alpha$-particle through a large barrier \cite{e}. A profound knowledge of this quantum mechanical effect enables one to obtain the Geiger-Nuttall law which relates the decay constant of a radioactive isotope with the energy of the $\alpha$ particles emitted.

WKB approximations have been applied to study $\alpha$-decay rate of many elements. Lifetimes of several heavy elements with $Z=102-120$ have been estimated by theoretically calculating the quantum mechanical tunneling probability in a WKB framework and also using the $DDM3Y$ effective nuclear interaction whose results have shown good agreement over a wide range of experimental datas \cite{aa,bb,cc,dd,ee,ff}.

Our acquaintance with the well known Geiger-Nuttall (GN) law is age old. In fact, the formulation of the Geiger-Nuttall (GN) law in 1911 was a landmark in itself. The GN law states that the $\alpha$-decay half life $T_{1/2}$ is related to the energy of $\alpha$-decay process ($\alpha$-decay value) $Q_{\alpha}$ as $log_{10}T_{1/2}=A(Z)Q_{\alpha}^{-1/2}+B(Z)$, where A(Z) and B(Z) are the coefficients which are determined by fitting experimental data \cite{f,g,h}. The GN law holds good for a restricted experimental data sets available but is invalid in general. In retrospect, we can see that although GN law gives a single straight line but if we consider the experimental data of $\alpha$-particle emitters including heavy and super heavy nuclei with proton numbers as large as 118, instead of getting a single linear path we observe several linear segments with different slopes and intercepts. This problem is overcome when the experimental results in logarithm form are plotted as a function of Viola-Seaborg (VS) parameter, $\tilde{V}=\frac{a^\prime Z_D+b^\prime}{\sqrt{Q_{\alpha}}}+c^\prime Z_D$, where $a^\prime$, $b^\prime$ and $c^\prime$ are constants \cite{i}. The striking feature of empirical Viola-Seaborg (VS) rule i.e $log_{10}T_{1/2}=\frac{a^\prime Z_D+b^\prime}{\sqrt Q_{\alpha}}+c^\prime Z_D+d^\prime$ with $a^\prime=1.478$, $b^\prime=8.714$, $c^\prime=-0.183$ and $d^\prime=-34.699$, is its superiority over GN law as it satisfactorily results in a straight line for a wide range of experimental data. As stated before the VS rule is precise but is an empirical one. Thus we strive for a formula of logarithm of half-lives in terms of well-defined parameters or coefficients so that the empirical nature of VS rule is apparently obscured.

Qi et al. have mentioned the validity and generalization of the GN law along with its microscopic basis. Also they have incorporated the tunneling process in the R-matrix theory and put forth an extended form of GN law\cite{j,k,l}.

To fully understand the R-matrix theory for the decay of a cluster or a particle, we look from the perspective of S-matrix theory of resonance scattering or the transition scattering from an isolated quasi-bound state to a scattering state as detailed in \cite{xyz}. For completeness, we can highlight the method as follows:

In S-matrix method, resonance is considered as a pole in the complex energy plane. Adding to that, the real part of the pole signifies the resonance energy or the Q-value of decay and the imaginary part represents the width which in turn gives the decay half-life of the system constituting of an $\alpha$-cluster and the residual nucleus \cite{bas}.

In the context of transition from quasi-bound state to a scattering state, we can write the width in terms of wave function at resonance and Coulomb functions in two ways:

i)By matching the normalized regular solution u(r) of the modified Schr\"{o}dinger equation and the distorted outgoing Coulomb function at a distance $r=R$ is
\begin{equation}
u(R)=N_0|G_0(\eta,kR)+iF_0(\eta,kR)|,
\end{equation}
where $F_0$ and $G_0$ are the regular and irregular Coulomb functions.
Thus, the decay width \cite{dav} is expressed as
\begin{equation}
\Gamma=\frac{\hbar^2k}{\mu}|N_0|.
\end{equation}

ii)We express the general formula of the $\alpha$-decay width \cite{lov} as follows:
\begin{equation}
\Gamma=2\pi|\langle\psi|H-H_0|\phi\rangle|^2,
\end{equation}
where $\psi$ is a bound initial state for the decaying nucleus and $\phi$ is a final scattering state for the $\alpha+$daughter system. The Hamiltonians $H_0$ and H are associated with $\phi$ and $\psi$, respectively.

Both the expressions are same in one way or another. Precisely in the first expression $\Gamma$ depends on a radial distance R which is represented crudely. This results in uncertainty in measuring certain data. It is interesting at this point to consider the second formula which is more effective in deriving an analytical expression for half-life in terms of the resonant wave function of an exactly solvable potential and the regular Coulomb function. Ultimately, by applying approximations on the functions of the expression of half-life, we derive a condensed formula for the logarithm of half-life in terms of the decay energy and mass and charge numbers of the $\alpha$-emitter. It is noteworthy that the found out condensed formula bear close resemblance with the VS rule. Thus we can say that finding half-lives from a derived formula is the paramount of this paper.

In section II we describe in detail the formulation and derivation of half-lives. In section III the formula is being implemented. Section IV includes the conclusion of the sought problem.


\section{Theoretical framework}
\subsection{Decay width or half-life of $\alpha$-decay}
We remark that the $\alpha$-decay process where the $\alpha$-cluster in the decaying nucleus is controlled by an attractive nuclear potential, $V_N(r)$ and the $\alpha$-particle outside the nucleus by the point-charge Coulomb potential i.e $V_C^P=\frac{Z_1Z_2e^2}{r}$.

In a simple picture we represent $H-H_0$ as the difference between the potentials in the two cases viz. the nuclear potential and the point-charge Coulomb potential i.e
\begin{equation}
H-H_0=\lbrace V_N(r)+V_C(r)\rbrace-V_C^P(r)=V_{eff}(r)-V_C^P(r),
\end{equation}
where $V_C(r)$ is the Coulomb potential given by 
\begin{equation}
V_C(r)=
\left\{
\begin{array}{cl}
0
&{if\;\;\;\;\; r > R_0,}\\
\frac{Z_1Z_2e^2}{2R_C}\lbrack3-(\frac{r}{R_C})^2\rbrack &{if\;\;\;\;\; r \le R_0,}
\end{array}
\right .
\end{equation}

In our approach, we calculate the decay width by taking into account the $\alpha$-decay process where there is transition of an $\alpha$-cluster from an isolated quasi-bound state to a scattering state. The initial system is related with the instability with the quasi-bound state of the decaying nucleus. Along with that the final state is the scattering state of the $\alpha$-daughter system.

Now, we solve the Schr\"{o}dinger equation using the effective potential which is the amalgamation of the nuclear potential and the electrostatic potential to get the radial part of the initial and final state of the wave function.

The radial part of the initial state wave function is
\begin{equation}
\psi_{nl}(r)=\frac{u_{nl}}{r}.
\end{equation}
Additionally, the final state wave function can be written considering the motion of the $\alpha$-particle relative to the daughter nucleus as a scattering state wave function corresponding to the $\alpha$-particle in point charge Coulomb potential \cite{fur}:
\begin{equation}
\phi(r)=\sqrt \frac{2\mu}{\pi \hbar^2 k}\hspace{2mm}\frac{F_l}{r}.
\end{equation}
where $k=\sqrt{2\mu E_{c.m.}/\hbar}$, $E_{c.m.}$ stands for the center-of-mass energy, $\mu=m_n\frac{A_{\alpha}A_D}{A_{\alpha}+A_D}$ is the reduced mass of the system with $m_n$ giving the mass of a nucleon, $A_\alpha$ represent the mass number of $\alpha$ particle, $A_D$ represent the mass number of the daughter nucleus and $F_l$ is the regular Coulomb wave function for a given partial wave $l$. The factor $\sqrt{\frac{2\mu}{\pi \hbar^2k}}$ is a normalization factor of the scattering wave function.

The mean-field approximation \cite{mah} is applied for the nucleon-nucleon interaction. Moreover, the process of double folding of the $\alpha+$nucleus potential $V_N(r)$ along with the electrostatic term, we get a parabola at a close radial distance r. Again simulation of the effective potential as a function of distance is done and is solved exactly in the Schr\"{o}dinger equation. Consequently the solution of which is the wave function $u_{nl}(r)$ for $l=0$ in the interior region.

Based on the Gell-Mann-Goldberger transformation \cite{dav}, the expression for the decay width becomes
\begin{equation}
\Gamma=\frac{4\mu}{\hbar^2k}\hspace{2mm}\frac{|\int_0^RF_l\lbrack V_{eff}(r)-V_C^P(r)\rbrack u_{nl}(r)dr|^2}{\int_0^R|u_{nl}(r)|^2dr}.
\end{equation}
For the normalization of the interior wave function the factor $\int_0^R|u_{nl}(r)|^2dr$ is used. The resonant wave function $u_{nl}(r)$ decreases rapidly with distance outside the Coulomb barrier radius $r_B$. For this reason, we apply the box normalization condition for the wave function $\int_0^R|u_{nl}(r)|^2dr=1$ for $R\approx r_B$.

Also, we are quite familiar with the relation between decay half-life $T_{1/2}$ and the width:
\begin{equation}
T_{1/2}=\frac{\hbar ln2}{\Gamma}.
\end{equation}
By using (9), we get a new expression of $T_{1/2}$
\begin{equation}
T_{1/2}=\frac{0.693\hbar^3k}{4\mu}\hspace{2mm}\frac{1}{J},
\end{equation}
\begin{equation}
J=\vert\int_0^RF_l\lbrack V_{eff}(r)-V_C^P(r)\rbrack u_{nl}(r)dr\vert^2.
\end{equation}
Now, the regular Coulomb wave function $F_l(r)$ can be expressed as \cite{fro} 
\begin{equation}
F_l=A_l\rho^{l+1}f_l(\rho),
\end{equation}
where $\rho=kr$, Sommerfeld parameter $\eta=\frac{\mu}{\hbar^2}\hspace{2mm}\frac{Z_{\alpha}Z_De^2}{k}$,
\begin{equation}
f_l(\rho)=\int_0^\infty (1-\tanh^2\epsilon)^{l+1}\hspace{1mm}\cos(\rho \tanh\epsilon-2\eta\epsilon)\hspace{1mm}d\epsilon,
\end{equation}
\begin{equation}
A_l=\frac{\sqrt {1-\exp(-2\pi\eta)}}{2^l\lbrace 2\pi\eta(1+\eta^2)(2^2+\eta^2)\ldots(l^2+\eta^2)\rbrace^{1/2}}.
\end{equation}
In particular, for $l=0$, $A_l$ is given by
\begin{equation}
A_0=\left\{\frac{1-\exp(-2\pi\eta)}{2\pi\eta}\right\}^{\frac{1}{2}}.
\end{equation}
We make a lay out (Figs. 1(a)-1(d)) to properly describe the radial dependence of the three terms in the integrand $J_1=\int_0^RF_l\lbrack V_{eff}(r)-V_C^P(r)\rbrack u_{nl}(r)dr$ mentioned in (11), the modulous of the resonance state wave function, $|u(r)|$, the regular Coulomb wave function, $F_0(r)$ for $l=0$ and the combined nuclear and Coulomb potential, $V_{eff}(r)$ by taking the $\alpha$+daughter system ($\alpha+\hspace{0.5mm}_{88}^{220}$Ra) with Q-value of decay or energy $E=7.298$ MeV representing the upper curve and the difference of potentials, $V_{eff}(r)-V_c^p(r)$ representing the lower curve of Fig. 1(c). In Fig. 1(d) the total integrand $J_1$ multiplied by a scaling factor of $10^{11}$ is shown. It is visible from the plot that the integrand shows a peak in the region close to the Coulomb barrier radius, $r_B=10.11$ fm. The plots indicate that the wave function of the resonance state decreases exponentially in the barrier region. Also the potential difference ie. $(V_{eff}-V_c^p)$ becomes zero in the region $r\ge r_B$. But the Coulomb function is very small at small values of r. We also find that, although the integrand J is dependent on distance r=R but it is independent of the distance R in the region $r\ge r_B$. Thus the value of decay time $T_{1/2}$ which is dependent on J is rather independent of R in the region $r\ge r_B$.
\subsection{The effective $\alpha$+nucleus potential}
 We use a potential \cite{fie} which is a function of radial variable r and can be represented in the form as follows:
\begin{equation}
V_{eff}(r)=
\left\{
\begin{array}{cl}
V_0[S_1+(S_2-S_1)\rho _1]
&{if\;\;\;\;\; r \le R_0,}\\
V_0S_2\rho _2 &{if\;\;\;\;\; r \ge R_0,}
\end{array}
\right .
\end{equation}
where $V_0$ is the strength of the potential with value $1$ MeV.
\begin{equation}
\rho_n=\frac{1}{\cosh^2\frac{R_0-r}{d_n}};n=1,2,\nonumber\\
\end{equation}
$d_n$ accounts for the flatness of the barrier, $d_1$ deciding the steepness of the interior side of the barrier whereas the exterior side is judged by $d_2$. $R_0$ is the radial position having value; $R_{0}=r_{0}(A_\alpha ^{1/3}+A_D^{1/3})+2.72$, $r_{0}=0.97$ fm.  Since we are considering the $\alpha$+nucleus system, $Z_\alpha$ represent the proton number of $\alpha$ particle, $Z_D$ represent the proton number of the daughter nucleus. Moreover $S_1$ and $S_2$ are the depth and height of the potential, respectively, having values;
\begin{equation}
S_1=-78.75+\frac{3Z_\alpha Z_D e^2}{2R_c},\nonumber \\ 
\end{equation}
\begin{equation}
S_2=\frac{Z_\alpha Z_D e^2}{R_0}(1-\frac{a_g}{R_0}),\nonumber\\
\end{equation}
where $R_c$ is the Coulomb radius parameter; $R_c=r_c(A_\alpha^{1/3}+A_D^{1/3})$, $a_g=1.6$ fm, $r_c=1.2$ fm, $e^2=1.43996$ MeV fm. $r_c$ and $a_g$ are the distance parameters.

\begin{figure}
\includegraphics[width=1.0\columnwidth,clip=true]{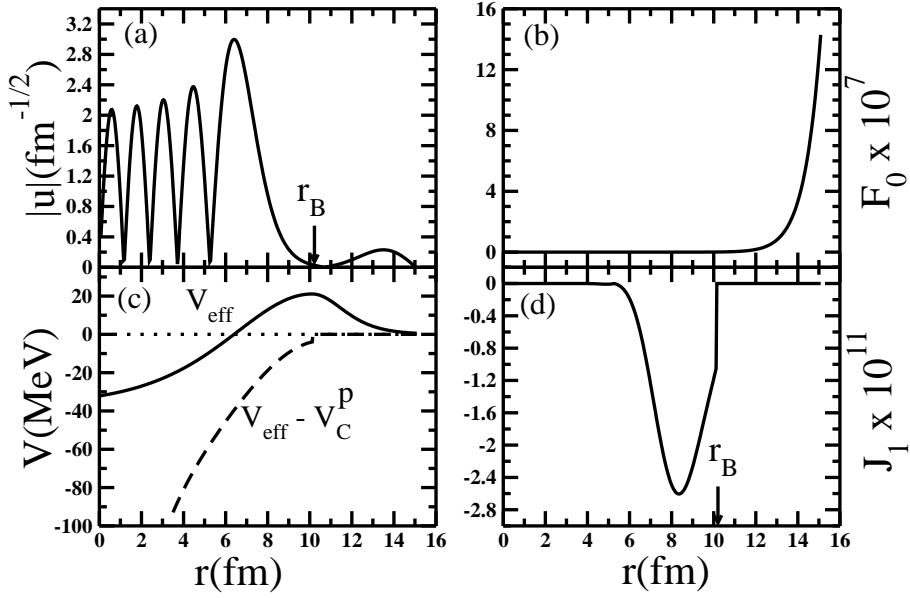}
\label{Fig.1}\vspace{0.2cm} \caption{Various terms for explaining the $\alpha$-decay rate in s-wave of $\alpha+_{88}^{220}$Ra system: (a) the modulous of the radial wave function at resonance, (b) the regular Coulomb wave function multiplied by $10^7$, (c) the $\alpha+$ daughter potential, $V_{eff}$ representing the upper plot and $V_{eff} -V_C^P$ representing the lower plot in dotted line, (d) the integral $J_1=\int_0^RF_l\lbrack V_{eff}(r)-V_C^P(r)\rbrack u_{nl}(r)dr$ multiplied by $10^{11}$ for the s-wave. The barrier radius $r_B=10.11$ fm is shown in arrows.}\label{figOne.}
\end{figure}

\subsection{Expression for decay half-life}
As previously mentioned, we are considering the problem of $\alpha+$nucleus system with a specific energy value $Q_{\alpha}$ value and radius $R=r_B$, the values of Sommerfeld parameter $\eta$ and parameter $\rho=kR$ are such that $\eta\rho\le50$ and $\rho\approx 10$. In this context, we now use the power series expansion and write the Coulomb wave function $F_l(r)$ as
\begin{equation}
F_l^{ps}(r)=C_l\rho^{l+1}G_l,
\end{equation}
\begin{equation}
(n+1)(n+2l+2)G_{n+1}=2\eta\rho G_n-\rho^2G_{n-1},
\end{equation}
\begin{equation}
G_0=1, G_1=\frac{\eta\rho}{(l+1)}, G_l=\sum_j G_j,
\end{equation}
\begin{equation}
C_l^2=\frac{P_l(\eta)}{2\eta}\frac{C_0^2(\eta)}{(2l+1)},
\end{equation}
\begin{equation}
P_l(\eta)=\frac{2\eta(1+\eta^2)(4+\eta^2)\ldots(l^2+\eta^2)2^{2l}}{(2l+1)\lbrack(2l)!\rbrack^2}.
\end{equation}
For the case, $l=0$, $P_0=2\eta$, and
\begin{equation}
C_0^2(\eta)=2\pi \eta\lbrace\exp(2\pi\eta)-1\rbrace^{-1},
\end{equation}
By putting the value of $G_0$
\begin{equation}
F_0^{ps}(r)=C_0\rho G_0,
\end{equation}
We find that
\begin{equation}
F_l(r)=x_mF_l^{ps}(r),
\end{equation}
where $x_m\approx70$. Therefore, instead of computing function $F_l(r)$ using (12), we go by the simple power series expansion of $F_l^{ps}$ using (20) multiplied by a factor $x_m=70$. Fig. 1 clearly show that the magnitude of this function is zero near the origin $r=0$ but increases predominantly at $r=r_B$ whereas the resonant wave function $u_{nl}(r)$ is very small beyond $r=r_B$. Hence, the integral J can be written in terms of $F_l(r)=x_mF_l^{ps}(r)$ at a point $r=R=r_B$ alongwith some multiplying factor which take care of the other contributions within the region $0<r<R$.

The integral J now changes to
\begin{equation}
J=|c_fF_l(R)|^2=|c_fx_mF_l^{ps}(R)|^2.
\end{equation}
For a typical $\alpha+$nucleus system, the value of $c_f(=\sqrt J/|x_m F_l^{ps}(R)|)$ is found to be $0.8$ for $l=0$.
Using the above found J value, the decay half-life $T_{1/2}$ becomes
\begin{equation}
T_{1/2}=0.693\hbar q_l\frac{\exp(2\pi\eta)}{f_{ml}^2},
\end{equation}
where
\begin{equation}
q_l=\frac{2\eta(2l+1)}{P_l(\eta)\rho^{2l}},
\end{equation}
\begin{equation}
f_{ml}=\sqrt{4\mu/\hbar^2k} \sqrt{2\pi\eta}c_fx_m\rho G_l,
\end{equation}
We take the logarithm of both sides,
\begin{equation}
logT_{1/2}=a\chi+c+d+b_l,
\end{equation}
\begin{equation}
a=1.4398\pi\sqrt{2(931.5)}/197.329
\end{equation}
\begin{equation}
\chi=Z_{\alpha}Z_D\sqrt{\frac{A_{\alpha}A_D}{(A_{\alpha}+A_D)Q_{\alpha}}},
\end{equation}
\begin{equation}
c=-2logS,
\end{equation}
\begin{equation}
d=-2logD,
\end{equation}
\begin{equation}
b_l=log(q_l),
\end{equation}
\begin{equation}
S=c_fx_mRG_l\frac{A_{\alpha}A_D\sqrt{Z_{\alpha}Z_D}}{A_{\alpha}+A_D},
\end{equation}
\begin{equation}
D=\frac{2\times931.5\times\sqrt{1.4398\times2\pi}}{(197.329)^2\sqrt{0.693\times197.329\times0.333\times10^{-23}}}.
\end{equation}

The expression (32) is some what similar to the VS relation mentioned previously but the difference is that in the present case the parameters and coefficients namely $a$, $c$, $d$, $b_l$ are well defined. The value of the coefficient $a=0.96$ and constant $d=-45.262$. The parameter c (35) depends on $Z_D$, $A_D$, $Q_{\alpha}$ and angular momentum partial wave $l$. Further, the parameter $b_l$ is related to $q_l$ (37) specifying the decay time for an $\alpha$ particle emitting with different angular momentum $l$.
\begin{figure}
\includegraphics[width=1.0\columnwidth,clip=true]{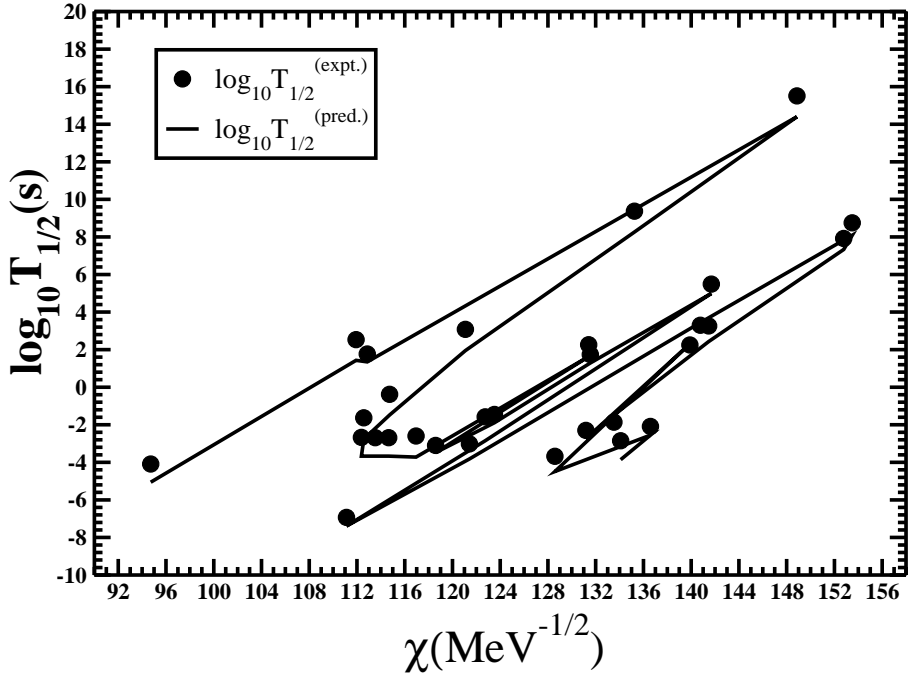}
\label{Fig.1}\vspace{0.2cm} \caption{Plot of decimal logarithm of half-lives $log_{10}T_{1/2}^{(expt.)}$ from experiments (solid dots) \cite{wan,aud} and $log_{10}T_{1/2}^{(pred.)}$ from calculation using (32) (solid line) as a function of $\chi $ in $l=0$ state for $\alpha$ emitters with $Z=52-118$.}\label{figOne.}
\end{figure}

\begin{figure}
\includegraphics[width=1.0\columnwidth,clip=true]{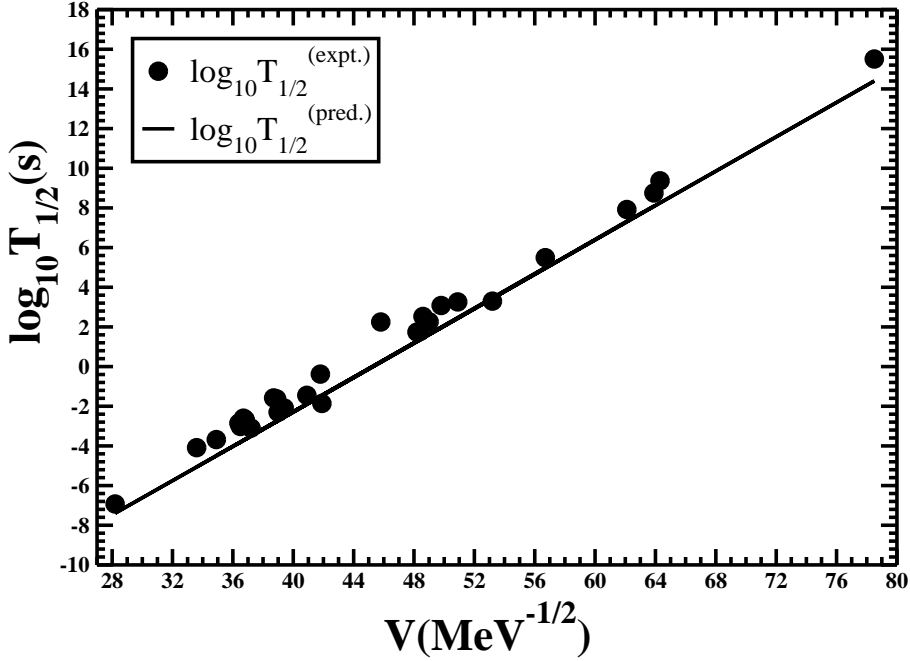}
\label{Fig.1}\vspace{0.2cm} \caption{Plot of decimal logarithm of half-lives $log_{10}T_{1/2}^{(expt.)}$ from experiments (solid dots) \cite{wan,aud} and $log_{10}T_{1/2}^{(pred.)}$ from calculation using (32) (solid line) as a function of $V=a\chi +c$ in $l=0$ state for $\alpha$ emitters with $Z=52-118$.}\label{figOne.}
\end{figure}

\section{Results and Discussions}
We take the solvable potential described in Eqn.(16) and denote it as effective Coulomb-nuclear potential for the $\alpha+$nucleus system and change only the steepness of the interior side of the barrier i.e. $d_1$. With the $Q_\alpha$ value with us and the wave function $u_{nl}(r)$ at resonance, we calculate the half-life $T_{1/2}$ by using (10) and denote it by $T_{1/2}^{(calt.)}$.

For easy handling of the problem we condense the integral J given by (11) and write in terms of $c_f$, $x_m$ and $F_l^{ps}$ as mentioned in Eqn.(28). Our analysis show that for different nuclei the values of $c_f$ comes out to be in the range 0.4 to 1.5 for $l=0$. We assign the value of $c_f$ ie. $c_f=0.8$ for $l=0$ and $c_f=0.6$ for $l>0$. Furthermore, using this $c_f$, we estimate the values of $T_{1/2}$ by using the closed form expression (32) for the decimal logarithm of half-life and represent it as $log_{10} T_{1/2}^{(pred.)}=log T_{1/2}^{(pred.)}/2.30258$. At last we determine the values of $T_{1/2}^{(pred.)}$ from the already found out $log_{10}T_{1/2}^{(pred.)}$. We then compare the calculated results $T_{1/2}^{(calt.)}$ using (10), experimental results $T_{1/2}^{(expt.)}$ and predicted $T_{1/2}$ values i.e $T_{1/2}^{(pred.)}$ using (32) and present systematically in Table 1 for $l=0$ case. For a large assemblage of nuclei starting from Z=52 to 118, we write the $T_{1/2}$ values viz. $T_{1/2}^{(expt.)}$, $T_{1/2}^{(calt.)}$, $T_{1/2}^{(pred.)}$. Our findings reveal that we get a wide band of $T_{1/2}$ ranging from decimal logarithmic values of -7.39 s to 14.4 s.
In the same way, we write the $T_{1/2}^{(expt.)}$, $T_{1/2}^{(calt.)}$, $T_{1/2}^{(pred.)}$ for $l>0$ by taking $c_f=0.6$ and present in Table 2. Our findings show that by changing the effective potential the $c_f$ value is changing keeping in mind the results of Ref.\cite{xyz}.  

\begin{table*}
\caption{\label{tab:table1}Comparison of experimental results of ground-state to ground state $(l=0)$ $\alpha$-decay half-life $log_{10}T_{1/2}^{(expt.)}$ in seconds with the calculated results $log_{10}T_{1/2}^{(calt.)}$ in seconds using (10) and predicted values $log_{10}T_{1/2}^{(pred.)}$ in seconds using (32). The experimental data of $\alpha$-decay half-life and $Q_\alpha$ value in MeV are obtained from Ref.\cite{wan,aud}. The value of the parameter $d_2$ is kept constant throughout i.e. $d_2=2$.}
\renewcommand{\tabcolsep}{0.2cm}
\renewcommand{\arraystretch}{1.6}
\footnotesize
\begin{tabular}{c c c c c c}
\hline
\hline
Nucleus&$d_1$&Q$_\alpha^{expt.}$(MeV)& $log_{10}T_{1/2}^{expt.}(s)$& $log_{10}T_{1/2}^{calt.}(s)$&$log_{10}T_{1/2}^{pred.}(s)$
\\ \hline\hline
$^{106}_{52}$&6.41600& 4.290 & -4.09  & -4.47 & -5.06 
 \\
$^{112}_{54}$&4.77400& 3.330 & 2.53  & 1.80 & 1.43
  \\
$^{114}_{56}$&4.72599& 3.534&1.77 & 1.67 & 1.35 
 \\
$^{146}_{62}$&4.90799& 2.528& 15.51& 15.15 &14.4
 \\
$^{148}_{64}$&4.94400& 3.271 & 9.37 & 10.05 &8.26
  \\
$^{150}_{66}$&5.02799& 4.3510 & 3.08 & 3.80 &1.95
  \\
$^{154}_{70}$&5.04400& 5.474& -0.38& -0.40&-1.48
  \\
$^{156}_{72}$&6.50199& 6.028 & -1.63& -0.67 &-2.76
 \\
$^{162}_{76}$&5.03600& 6.767 & -2.67 & -2.34 &-3.67
 \\
$^{168}_{78}$&5.06200& 6.999 & -2.69& -2.94&-3.67
 \\
$^{174}_{80}$&5.09399& 7.233 &-2.69 & -2.11&-3.67
 \\
$^{190}_{84}$&5.21400& 7.699 & -2.60 & -2.61 &-3.72
 \\
$^{218}_{84}$&5.41800& 6.115 & 2.26 & 4.34&1.63
 \\
$^{194}_{86}$&5.18000& 7.862 & -3.10 & -1.58&-3.48
 \\
$^{218}_{86}$&4.22399& 7.263& -1.45 & -1.37&-1.89
 \\
$^{220}_{86}$&4.12200& 6.405 & 1.74  & 1.97 &1.29
 \\
$^{224}_{88}$&5.19999& 5.789 & 5.49  & 7.99 &4.99
   \\
$^{216}_{90}$&4.07800& 8.071 & -1.58& -2.37&-2.85
 \\
$^{218}_{90}$&4.39800& 9.849 & -6.93 & -7.06 &-7.39
 \\
$^{224}_{92}$&5.45800& 8.633 & -3.02& -1.92&-3.82
  \\
$^{232}_{94}$&6.48600& 6.716 & 3.30& 6.44&3.44
  \\
$^{244}_{96}$&5.03000& 5.902 & 8.75 & 11.13&8.09
  \\
$^{252}_{98}$&5.07000& 6.217 & 7.92 & 10.63 &7.33
 \\
$^{250}_{100}$&5.147990& 7.556 & 3.26 & 5.27 &2.43
 \\
$^{258}_{104}$&4.05800& 9.190& -1.86 & -0.26&-1.48
 \\
$^{260}_{106}$&5.34000& 9.901 & -2.30& -0.68&-2.74
 \\
$^{270}_{108}$&6.58800& 9.050 & 2.25& 3.01&2.31
 \\
$^{270}_{110}$&4.18599& 11.120 & -3.68 & -3.05 &-4.49
 \\
$^{290}_{116}$&4.07999& 10.990& -2.09& -0.77&-2.55
 \\
$^{294}_{118}$&4.15000& 11.810& -2.85& -2.22&-3.86
 \\

\hline\hline
\end{tabular}
\end{table*}

\begin{table*}
\caption{\label{tab:table1}Logarithm of calculated $\alpha$-decay half-lives $log_{10}T_{1/2}^{(calt.)}$ in seconds using (10), logarithm of predicted $\alpha$-decay half-lives $log_{10}T_{1/2}^{(pred.)}$ in seconds using (32) with parameter fixed $c_f=0.8$ for $l=0$ and $c_f=0.6$ for $l>0$. The experimental results of half-lives $log_{10}T_{1/2}^{(expt.)}$ in seconds \cite{aud}. The $\alpha$-decay energies $Q_{\alpha}$ in MeV are taken from atomic mass table \cite{wan}. The value of the parameter $d_2$ is kept constant throughout i.e. $d_2=2$.}
\renewcommand{\tabcolsep}{0.18cm}
\renewcommand{\arraystretch}{1.6}
\footnotesize
\begin{tabular}{c c c c c c c}
\hline
\hline
Nucleus&$d_1$&Q$_\alpha^{expt.}$(MeV)&$l$& $log_{10}T_{1/2}^{expt.}(s)$& $log_{10}T_{1/2}^{calt.}(s)$&$log_{10}T_{1/2}^{pred.}(s)$
\\ \hline\hline
$^{105}_{52}$&4.88800& 4.889 &0& -6.19  & -7.42 & -7.45 
 \\
$^{109}_{54}$&6.32800& 4.217 &2& -1.52  & -2.57 & -3.83
  \\
$^{147}_{62}$&4.89599& 2.331 &0& 18.52 & 17.05 & 14.50  
 \\
$^{157}_{70}$&4.96400& 4.621&0 & 1.58& 4.32 &2.60
 \\
$^{159}_{74}$&5.04800& 6.450 &0 & -2.08 & -2.74 &-3.43
  \\
$^{165}_{74}$&4.93400& 5.029 &2 & 0.70 & 4.47 &2.24
  \\
$^{161}_{76}$&5.06400& 7.066& 0& -3.19& -3.93&-4.61
  \\
$^{167}_{78}$&5.06599& 7.160 & 0& -3.09& -2.58 &-4.17
 \\
$^{171}_{78}$&5.05599& 6.610 & 0& -1.34 & -1.29 &-2.40
 \\
$^{175}_{80}$&5.07399& 7.043 & 0& -1.97& -1.62&-3.08
 \\
$^{187}_{84}$&6.65999& 7.979 &2& -2.85 & -2.37&-4.78
 \\
$^{189}_{84}$&5.19199& 7.701 & 2& -2.42 & -2.01 &-3.99
 \\
$^{209}_{84}$&5.07600& 4.979 &2&  9.50 & 8.93 &7.28
 \\
$^{211}_{84}$&4.28200& 7.595 &5 & -0.28 & -3.07&-3.69
 \\
$^{209}_{86}$&6.54600& 6.156 &0&  3.23 & 5.46&2.50
 \\
$^{211}_{86}$&5.14599& 5.965 &2&  4.72  & 6.18 &3.06
 \\
$^{213}_{86}$&4.29400& 8.244 & 5& -1.70  & -3.61 &-4.86
 \\
$^{221}_{86}$&4.09399& 6.148 &2&  3.18& 4.27&2.10
 \\
$^{203}_{88}$&5.19400& 7.730 & 0& -1.44 & -0.52 &-2.41
 \\
$^{217}_{90}$&4.30800& 9.433 &5& -3.60& -5.40& -6.54
  \\
$^{221}_{90}$&4.24200& 8.628 & 2& -2.77& -3.05&-4.79
  \\
$^{225}_{90}$&5.28600& 6.920 & 2& 2.72 & 3.64&6.59
  \\
$^{219}_{92}$&4.30600& 9.940 &5&  -4.25 & -6.59 &-7.06
 \\
$^{235}_{92}$&4.935990& 4.678 &1&  16.34 & 16.26 &13.3
 \\
$^{239}_{94}$&4.97000& 5.245& 3& 11.88 & 14.17&10.6
 \\
$^{243}_{96}$&5.05399& 6.169 & 2& 8.96& 9.95&6.45
 \\
$^{245}_{96}$&4.99200& 5.622 &2&  11.42& 12.91&9.35
 \\
$^{247}_{96}$&4.97599& 5.3540 &1&  14.69 & 14.64 &10.8
 \\
$^{249}_{98}$&5.04400& 6.296& 1& 11.4& 9.18&6.60
 \\
$^{251}_{98}$&5.05399& 6.177& 5& 10.45& 10.28&7.51
 \\
$^{251}_{100}$&6.52000& 7.425& 1& 4.28& 5.87&2.54
 \\
$^{253}_{100}$&5.13600& 7.199&5 & 5.41& 6.47&3.72
 \\
$^{255}_{100}$&5.17399& 7.240&4&  4.87& 6.35&3.36
 \\
$^{257}_{100}$&5.13800& 6.864& 2& 6.95& 8.10&4.86
 \\
$^{267}_{110}$&5.53399& 11.780& 0& -5.0& -3.85&-5.85
 \\
$^{273}_{110}$&4.26599& 11.370& 0& -3.61& -3.95&-5.07
 \\
$^{277}_{112}$&6.97000& 11.620& 0& -3.00& -2.73&-5.05
 \\
$^{291}_{116}$&4.07200& 10.8900& 0& -1.55& -1.19&-2.32
 \\

\hline\hline
\end{tabular}
\end{table*}
Now to clarify the ambiguity on the non linearity of G N law for various $\alpha$-emitters, we plot $log_{10}T_{1/2}^{(expt.)}$ and $log_{10}T_{1/2}^{(pred.)}$ as a function of the Coulomb parameter $\chi=Z_{\alpha}Z_{D}\sqrt{\frac{A_{\alpha}A_D}{(A_{\alpha}+A_D)Q_{\alpha}}}$ for $l=0$ and present the plot in Fig. 2. Likewise, we also make a plot of $log_{10}T_{1/2}^{(expt.)}$ and $log_{10}T_{1/2}^{(pred.)}$ as a function of $V=a\chi+c$ used in (32) and present in Fig. 3. To our utter surprise, we find that the plot of $log_{10}T_{1/2}^{(expt.)}$ and $log_{10}T_{1/2}^{(pred.)}$  vs $\chi$ gives multiple straight lines whereas $log_{10}T_{1/2}^{(expt.)}$ and $log_{10}T_{1/2}^{(pred.)}$  vs $V=a\chi+c$ gives a single straight line. This clearly indicate that the our measured results have great accuracy.

\section{Summary and Conclusion}
By using the regular Coulomb function, resonant wave function and the difference in potentials a general formula is being put forth for the calculation of $\alpha$-decay width. The $\alpha$+nucleus potential generated by using relativistic mean field theory is closely reproduced by special expressions of the potential and finally we zero in to a closed formula for the logarithm of half-life as a function of Q-values of various decaying nuclei of different masses and charges. This derived formula is impeccable in finding the logarithm of half-lives. The closed formula for the logarithm of half-life favorably explains the half-lives ranging from $10^{-6}$s to $10^{22}$y. Also this closed form expression curtains the dilemma over nonlinearity as it fairly reproduces the rectilinear alignment of the logarithm of the experimental decay half-lives as a function of the Viola-Seaborg parameter. With the updated VS rule with us, we only need the $Q_\alpha$ value, mass number $A_D$ and charge $Z_D$ to predict the $\alpha$-decay half-life and there by future work concerning half-lives of all types of nuclei can be foresighted.

\section{Acknowledgement}
We gratefully acknowledge the computing and library facilities extended by the Institute of Physics, Bhubaneswar.


\begin{thebibliography}{99}
\bibitem{a} M. A. Preston,{\textit{Phys. Rev.}} {\bf 71}, 865(1947).
\bibitem{b} I. Perlman, A. Ghiorso, and G. T. Seaborg,{\textit{Phys. Rev.}} {\bf 77}, 26(1950).
\bibitem{c} M. Balasubramaniam and N. Arunachalam, {\textit{Phys. Rev.}} {\bf C71}, 014603 (2005).
\bibitem{d} Y. Qian, Z. Ren, {\textit{ Phys. Lett.}}  {\bf B738}, 87 (2014). 
\bibitem{e} G. Gamow, {\textit{Z. Phys.}} {\bf51}, 204 (1928).
\bibitem{aa} D. N. Poenaru, I. H. Polonski, R. A. Gherghescu, and W. Greiner, {\textit{J. Phys. G:Nucl. Part. Phys.}} {\bf32}, 1223(2006).
\bibitem{bb} D. N. Basu, {\textit{J. Phys. G: Nucl. Part. Phys.}} {\bf30}, B35 (2004).
\bibitem{cc} C. Samanta, P. R. Chowdhury, and D. N. Basu, {\textit{Nucl. Phys.}} {\bf A789}, 142(2007).
\bibitem{dd} P. R. Chowdhury, C. Samanta, and D. N. Basu, {\textit{Phys. Rev.}} {\bf C73}, 014612(2006).
\bibitem{ee} M. Horoi, B. A. Brown, and A. Sandulescu, {\textit{J. Phys. G: Nucl. Part. Phys.}} {\bf30}, 945(2004).
\bibitem{ff} H. Hassanabadi, E. Javadimanesh, and S. Zarrinkamar, {\textit{Int. J. Mod. Phys.}} {\bf E22}, 1350080(2013).
\bibitem{f} C. Qi et al., {\textit{ Phys. Lett.}}  {\bf B734}, 203 (2014).
\bibitem{g} C. Qi et al.,{\textit{J. of Phys.:Conference series}} {\bf381}, 012131(2012).
\bibitem{h} Y. Ren and Z. Ren, {\textit{Phys. Rev.}} {\bf C85}, 044608 (2012).
\bibitem{i} D. S. Delion and A. Dumitrescu, {\textit{At. Data Nucl. Data Tables}} {\bf 101}, 1 (2015).
\bibitem{j} C. Qi, F. R. Xu, R. J. Liotta, and R. Wyss, {\textit{Phys. Rev. Lett.}} {\bf 103}, 072501 (2009).
\bibitem{k} C. Qi, F. R. Xu, R. J. Liotta, and R. Wyss, M. Y. Zhang, C. Asawatangtrakuldee, and D. Hu, {\textit{Phys. Rev.}} {\bf C80}, 044326 (2009).
\bibitem{l} C. Qi, A. N. Andreyev, M. Huyse, R. J. Liotta, P. Van Duppen, and R. Wyss, {\textit{Phys. Lett.}} {\bf B734}, 203 (2014).
\bibitem{xyz} B. Sahu and S. Bhoi, {\textit{Phys. Rev.}} {\bf C93}, 044301 (2016).
\bibitem{bas} B. Sahu, Y. K. Gambhir, and C. S. Shastry,{\textit{Mod. Phys. Lett.}} {\bf A25}, 535 (2010).
\bibitem{dav} C. N. Davids and H. Esbensen, {\textit{Phys. Rev. C}} {\bf C61}, 054302(2000).
\bibitem{lov} R. G. Lovas, R. J. Liotta, A. Insolia, K. Varga, and D. S. Delion, {\textit{Phys. Rep.}} {\bf 294}, 281 (1998).
\bibitem{fur} V. I. Furman, S. Holan, S. G. Kadmensky, and G. Stratan,  {\textit{Nucl. Phys.}} {\bf A226}, 131 (1974).
\bibitem{mah} S. Mahadevan, P. Prema, C. S. Shastry, and Y. K. Gambhir,  {\textit{Phys. Rev.}} {\bf C74}, 057601(2006).

\bibitem{fro} C. E. Fr\"{o}berg, {\textit{Rev. Mod. Phys.}} {\bf 27}, 399(1955).
\bibitem{fie} H. Fiedeldey, W. E. Frahn, {\textit{Annls.of Phys.}} {\bf16}, 387 (1961).
\bibitem{wan} M. Wang {\textit{et al.}}, {\textit{Chin. Phys.}} {\bf C36}, 1603 (2012).
\bibitem{aud} G. Audi {\textit{et al.}},  {\textit{Chin. Phys.}} {\bf 16}, 1157 (2012).
\end{thebibliography}
\end{document}